\documentstyle[12pt,epsfig]{article}

\newcommand {\bi} {\bibitem}
\newcommand {\be} {\begin{equation}}
\newcommand {\beq} {\begin{eqnarray} \nonumber }
\newcommand {\ee} {\end{equation}}

\newcommand{\bq}{\begin{eqnarray}}
\newcommand{\eq}{\end{eqnarray}}

\newcommand{\simg}{\stackrel{>}{\sim}}
\newcommand{\bc}{\begin{center}}
\newcommand{\ec}{\end{center}}

\begin{document}


\title{On the Energy Minima of the SK Model}
\author{Barbara Coluzzi$^{a}$, Enzo Marinari$^{a}$,
Giorgio Parisi$^{a}$ and Heiko Rieger$^{b}$}

\maketitle

\begin{center}

a) Dipartimento di Fisica and Sezione INFN,\\
Universit\`a di Roma ``La Sapienza'',
Piazzale Aldo Moro 2,
I-00185 Rome (Italy) \\

b) FB 10.1 Theoretische Physik \\   
   Universit\"at des Saarlandes, 66041 Saarbr\"ucken (Germany)         


\end{center}

\vspace{2cm}

\begin{abstract}
\noindent 
We study properties of the energy minima obtained by quenching equilibrium 
configurations of the Sherrington-Kirkpatrick (SK) mean field spin glass. 
We measure the probability distribution of the overlap among quenched 
configurations and the quenched energy, looking at the dependence on the 
starting equilibrium temperature, and performing a systematic analysis of 
finite size effects.
\end{abstract}


\newpage

\begin{section}{Introduction.}
\noindent
The local energy minima properties are analyzed in several works on
glass-forming liquids and they are allowing a better understanding of
the behavior of these systems. The importance of potential energy
landscape in the physics of super-cooled liquids was already pointed
out by Goldstein \cite{Go}. More recently Stillinger and Weber
\cite{StWe} formalized the idea that the multidimensional energy
surface can be partitioned in a large number of local minima, so
called {\it Inherent Structures} (IS), each one surrounded by its
attraction basin. It is now clear \cite{ScSaDyGl} that the
low-temperature dynamics (i.e. for temperatures below the Mode
Coupling critical temperature $T_{MCT}$ \cite{Goe}) can be subdivided
into intra-basin motion and crossing of energy barriers by activated
processes, taking place on a significantly longer time-scale.

The system at equilibrium below $T_{MCT}$ is `almost always' trapped
in one of the basins accessible at this temperature. The huge number
${\cal N} \propto \exp(N \Sigma)$ of these `valleys', exponentially
diverging with the system size $N$, suggested the scenario of an
underlying thermodynamic transition due to an `entropy crisis' at the
Kauzmann temperature $T_K<T_{MCT}$ where the configurational entropy
$\Sigma$ goes to zero \cite{Ka}, which was supported by recent
analytical work \cite{MePa}- \cite{CoMePaVe}. By looking at IS one can
evaluate $\Sigma$ numerically \cite{CoMePaVe}-\cite{BuHe}. Moreover,
the IS energy turns out to be an interesting quantity for studying
both the static and the dynamical behavior
\cite{ScKoTa}-\cite{KoScTa}. Differences between fragile and strong
glasses were recently proposed to be explainable within an energy
landscape description, too \cite{Ca}.

The outlined picture of glass-forming liquids is reminiscent of that
characterizing generalized mean field spin glass models like  those
involving $p$-spin interactions \cite{CrSo}, which display a dynamical
ergodicity breaking at the temperature $T_D \equiv T_{MCT}$
\cite{CuKu1} (in this case the barriers between basins are infinite in
the thermodynamic limit also for $T_K < T < T_{MCT}$ because of the
mean field approximation), and a thermodynamic entropy driven
transition at a lower temperature $T_K$, corresponding to a one step
replica symmetry breaking (1RSB) scenario. For $T<T_K$ one finds a non
trivial probability distribution of the overlap between states $P(q)=m
\delta(q)+(1-m)\delta(q-q_{EA})$. Here $q_{EA}$ is the self-overlap of
a state with itself, whereas different states are orthogonal and have
mutual overlap zero.

Several years ago, Kirkpatrick, Thirumalai and Wolynes \cite{KiTh}
suggested that 1RSB spin glass models could be a paradigm of vitreous
systems. The numerical study of out-of-equilibrium dynamics in
glass-forming liquids gives intriguing results \cite{Pa1}-\cite{WaRi}.
The measurement of $P(q)$ among `glassy states' is a subtle task
\cite{CoPa,BhBrKrZi,DaVa}, since one faces both the problem of
thermalizing the system down to very low temperatures and that of
avoiding possible crystalline minima whose basin of attraction could
be non-negligible for small systems. As recently proposed by
Bhattacharya, Broderix, Kree and Zippelius \cite{BhBrKrZi}, to look at
the Inherent Structures is helpful also from this point of view, since
it allows a more precise definition of the overlap and make easier to
distinguish between glassy minima and crystalline or quasi-crystalline
configurations.

On the other hand, it is not completely clear $a~priori$ which kind of
behavior one should expect for the $P_{quen}(q)$ measured among energy
minima obtained by quenching equilibrium configurations instead of
among equilibrium configurations themselves. To our knowledge, such a
quantity was never previously studied in a spin glass model (which is
not surprising, since in this case equilibration is still feasible without 
difficulties for moderate system sizes). More generally, little is known 
about properties
of `Inherent Structures' in spin glasses, which (apart from analogies
with glass-forming liquids) have their own interest even in the well
understood Sherrington Kirkpatrick (SK) mean field model. In a
previous work \cite{Pa} one of the authors studied the energy minima of
the SK model obtained starting from random initial configurations
(i.e. by quenching from infinite initial temperature) whereas recently
Crisanti and Ritort \cite{CrRi,CrRi2} have performed both for a 1RSB
spin glass and for the SK model a numerical analysis of the static and
dynamical properties within the energy landscape description similar
to the one proposed in \cite{ScKoTa,KoScTa} for a glass-forming
liquid, by looking in particular at the IS energy and at the
configurational entropy. Their results confirm the close similarities
between 1RSB spin glass and structural glass energy landscapes and the
usefulness of this kind of approach, further suggesting a systematic
analysis.

In the present work, we perform an extensive numerical study of energy minima
properties in the SK model, considering initial equilibrium configurations 
both in the high-temperature (paramagnetic) region and deep in the glassy 
phase. We look first of all at the behavior of the appropriately defined 
$P_{quen}(q)$ which is compared with the corresponding (usual) equilibrium 
one. Then we extend the analysis to the overlap of the IS 
with the configurations from which they are obtained, which measure the 
(quite strong) correlations between them, and we study systematically finite 
size effects on the behavior of the IS energy.
\end{section}

\section{Model, Observables and Simulations.}
\noindent
The Sherrington Kirkpatrick spin glass model \cite{MePaVi} is described 
by the Hamiltonian
\be
{\cal H}_{J}=\sum_{i < j=1}^{N} J_{ij} \sigma_i \sigma_j\;,
\ee
where $\sigma_i=\pm 1$ are Ising spins, the sum runs over all pairs of
spins and $J_{ij}$ are random independent variables with mean value
$\overline{J_{ij}}=0$ and variance $1/N$. We take $J_{ij}=\pm
N^{-1/2}$.

This model is exactly solvable (since interactions have infinite
range) and has a glassy phase with full replica symmetry breaking
(FRSB). In case of zero magnetic field (the one we consider here),
taking into account the symmetry under inversion of the spins, the
$P(q)$ changes at the critical temperature $T_C=1$ from a
$\delta$-function at $q=0$ (characteristic of the paramagnetic phase)
to the FRSB two $\delta$-functions in $q=\pm q_{EA}$ with a non-zero
$plateau$ joining them. The transition is continuous also in the
order parameter (at variance with 1RSB models), i.e.  $\lim_{T
  \rightarrow T_C^{-}} q_{EA}(T)=0$, and there is no distinction
between the dynamical and the static transition \cite{CuKu2}, i.e.
$T_D = T_C$.

The SK model is particularly suitable for the kind of study we are interested
in, since its behavior is well understood and, on the other hand, by using
optimized Monte Carlo methods \cite{Ma} one is able to thermalize 
large system sizes down to low temperatures, which allows to study finite
size effects systematically. We simulated $N=64$, 128, 256, 512 and 1024, 
averaging over 
2048, 1024, 512, 384 and 192 different disorder realizations respectively. 
The program was multi-spin coded on different sites of the system (we store 64 
spins in the same word) and we used Parallel Tempering (PT) 
\cite{Ma,TeReOrWh}, running simultaneously two independent sets of copies 
(replica) of the system for each sample. Up to 50 (for the two 
largest sizes) different temperatures between $T_{min}=0.65$ and 
$T_{max}=3$ were used, and we performed from 100.000 PT steps for 
the smallest value of $N=64$ to 300.000 for the largest value of $N=1024$. 

The PT acceptance for the exchange 
of nearest neighbor temperatures was never smaller than 0.6. 
In the second half of the run we computed the specific heat
both as the derivative of the energy density with respect to temperature 
$c \equiv d \langle e \rangle/ dT$, and from fluctuations $c \equiv 
N(\langle e^2 \rangle- \langle e
\rangle^2)/T^2$, looking for compatibility of results. 
This means to compare one-time and two-time quantities
respectively, which is an effective way for checking thermalization
particularly when using PT (note that in this case
fluctuations involve different replicas evolving at the same
temperature at different times during the run). Nevertheless we also
divided the second half of the run in (four) equal intervals, checking
that there were no evident differences in the values of the considered
observables, $P(q)$ and $P_{quen}(q)$ in particular. A further confirm
for the system being well thermalized comes from the perfectly symmetric
with respect to the exchange $q \rightarrow -q$ probability distributions
that we obtained.

For each disorder realization and for each temperature, in the second
half of the run, we saved 1500+1500 equilibrium configurations from
the two independent sets of replica which were subsequently quenched
by a zero-temperature dynamics. The observables were computed from
these configurations and the corresponding energy minima, errors being
evaluated from sample-to-sample fluctuations. One should note that
both $P(q)$ and $P_{quen}(q)$ are strongly not self-averaging in the
glassy phase, wherefore it was necessary to average over a large
number of samples even for large system sizes. The whole simulations
would have taken about two years of CPU on a usual alpha-station, i.e.
a few days when using 128 processors simultaneously on Cray T3E (the
code is easily parallelized with efficiency close to 1 by running a
different disorder realization on each processor).

A subtle point concerns the quenching procedure. 
We are considering even values for $N$, which means that the local
field acting on each spin because of the other ones can never be zero.
Moreover, it is known \cite{Pa}, from the analysis of properties of
quenched configurations obtained starting from infinite temperature
(i.e.\ random initial configuration), that different zero temperature
dynamics give qualitatively identical results. The `Greedy algorithm',
where the spin corresponding to the largest energy decreasing is
flipped at each step, seems to stop more frequently in local high
energy minima than the `Reluctant algorithm', where one flips the spin
which gives the smallest energy decreasing (i.e., the contrary of the
previous case).  We choose to use an `intermediate' zero-temperature
dynamics, which is easy to implement too. At each step, a randomly
chosen spin is suggested to flip and at least $20 N$ steps are
performed after the last successful one before stopping. The
probability that the final configuration is not a local energy minimum
(under single spin flip) is therefore $\sim e^{-20}$, practically
negligible. As a last remark, it should be stressed that such a
quenching procedure could possibly give energy minima which are not
the `nearest' IS to the starting equilibrium configurations, i.e.
which are $less$ correlated but not $more$ correlated than in the
analog glass-forming liquid case, where one generates ISs by following
the path of steepest descent.

Labeling by $\{ \sigma_i \}$, $\{ \tau_i \}$ the spins belonging to two 
configurations, the overlap is defined as 
\be
{\cal Q}={1 \over N} \sum_{i=1}^N \sigma_{i} \tau_{i}.
\ee
The spin glass order parameter, i.e. the equilibrium probability distribution 
of overlap among states $P(q,T)$, is usually evaluated numerically as
the histogram of the instantaneous overlap ${\cal Q}$ between two replica 
(with the same disorder configurations) which evolve simultaneously and 
independently at temperature $T$
\be
P(q,T)=\overline{P_J(q)}=\overline{ \langle \delta(q-{\cal Q})\rangle}
\hspace{.1in},
\ee
where the thermal average $\langle \cdot \rangle$ corresponds to average 
over time in the simulation whereas $\overline{(\cdot )}$ stands for the 
average over $J_{ij}$ realizations. $P_J(q,T)$ can be equivalently measured 
from two given sets of equilibrium configurations belonging to the two
replica, by considering the overlap of each configuration of one set with
all the configurations of the other. In this work we evaluated $P_J(q,T)$
both during the simulation and from the saved configurations, obtaining 
perfectly compatible results, which confirm that these configurations sample
accurately enough the phase space.

We define the quenched probability distribution of the overlap as
\be
P_{quen}(q,T)=\overline{{1 \over {\cal N}^2_{IS}} 
\sum_{i_a,i_b=1}^{{\cal N}_{IS}} 
\delta(q-{\cal Q}_{IS})}\hspace{.1in},
\ee
where the sum runs over the ${\cal N}_{IS}=1500$ energy minima obtained
starting from the equilibrium configurations at temperature $T$ for
each of the two replica sets. This definition, analogous to the one
introduced in \cite{BhBrKrZi} for a Lennard-Jones glass-forming binary
mixture, implies that we are weighting each IS with the Boltzmann
factor of the corresponding basin at temperature $T$, as usual in
numerical studies on super-cooled liquids.

\section{Results.}
\subsection{The behavior of $P_{quen}(q)$.}
\begin{figure}[htbp]
\begin{center}
\leavevmode
\epsfig{figure=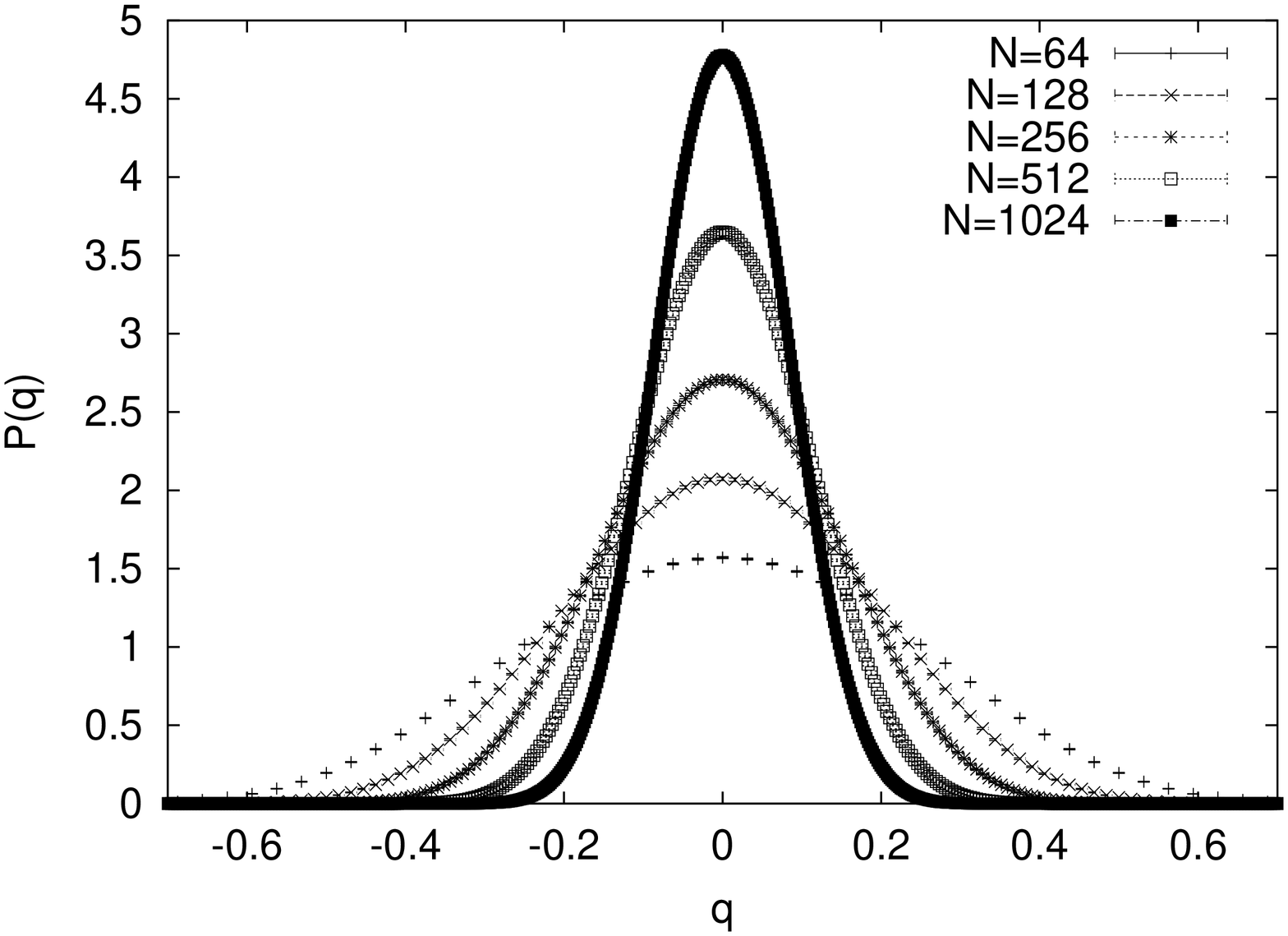,angle=270,width=8cm}
\epsfig{figure=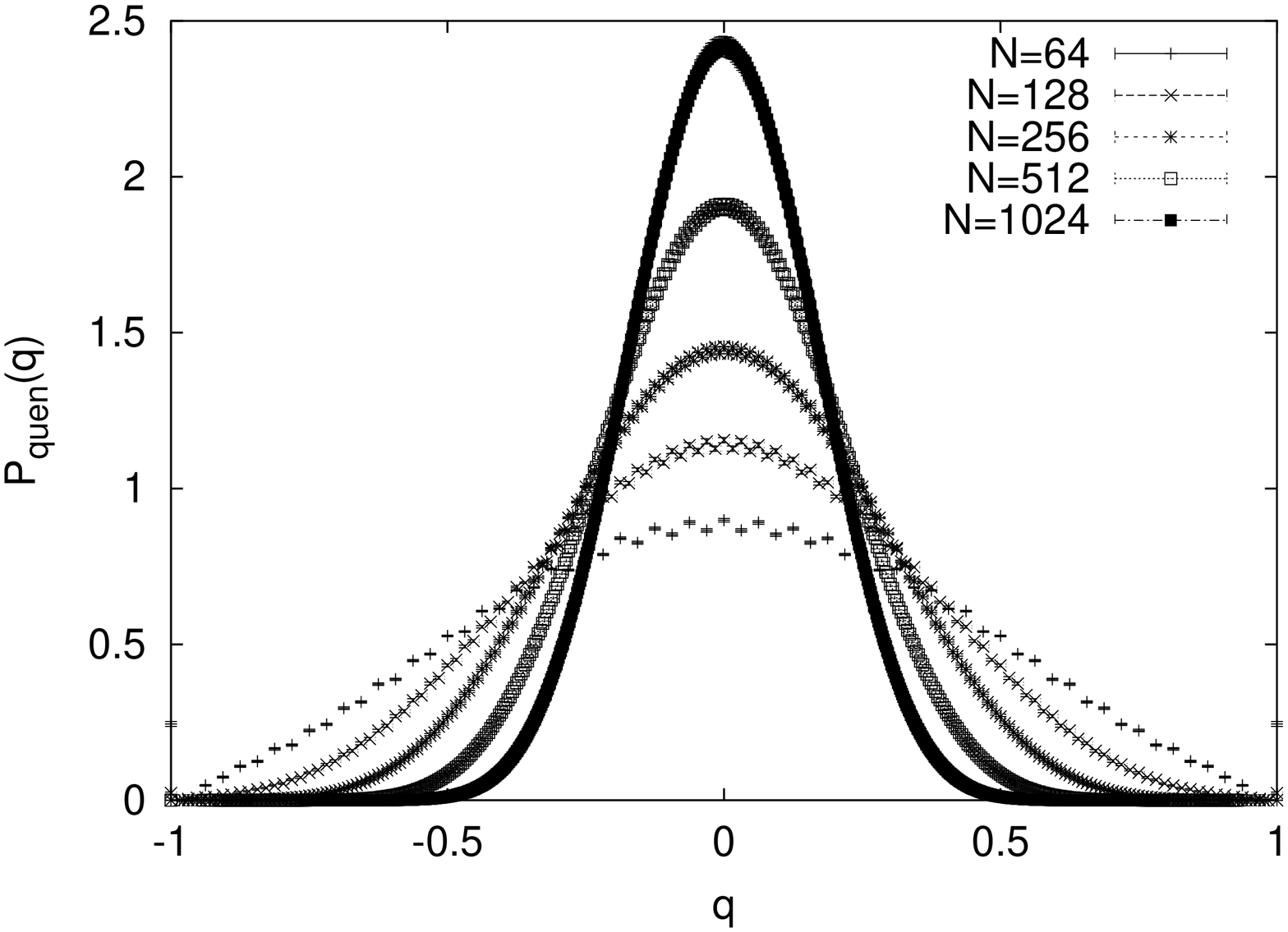,angle=270,width=8cm}
\caption{The equilibrium $P(q)$ (on the left) and the quenched one (on the 
right) at $T=1.05$, a value of the temperature slightly higher than $T_C=1$, 
for different values of the system size.}
\label{pqtc}
\end{center}
\end{figure}

We present in [Fig. \ref{pqtc}] data for the equilibrium overlap
distribution $P(q)$ (on the left) and for the quenched overlap
distribution $P_{quen}(q)$ (on the right) at a temperature very close
to the critical temperature $T=1.05=1.05~T_C$, but still in the paramagnetic
phase, for different system sizes. It was shown in \cite{Pa} that the
$P_{quen}(q)$ obtained from infinite temperature configurations
becomes more and more concentrated in $q=0$ for increasing $N$ and it
goes to a $\delta$-function in the thermodynamic limit. We find the
same qualitative behavior in the whole high-temperature region
$T>T_C=1$. It is crystal clear from [Fig. \ref{pqtc}] that there is no
evidence for replica symmetry breaking in the probability distribution
of overlap between IS reachable from equilibrium configurations at $T
\simg T_C$ and weighted with the Boltzmann factor of the corresponding
basin at this temperature.

In other words, we know that for $T<T_C$ there is RSB, but we cannot
detect it by looking at the probability distribution of the overlap
obtained with a fast quench starting from $T \simg T_C$.  This result
implies that also in the case of glass-forming liquid the analogously
defined $P_{quen}(q,T)$ will be trivial when quenching from the 
paramagnetic (liquid) phase, even when quenching down to a $T$ value
where RSB occurs (if it does). This is in agreement with the behavior
reported in \cite{BhBrKrZi} for a Lennard-Jones binary mixture starting
from $T \simg T_{MCT}$.

In glass-forming liquids the overlap among IS is easier to define than
the equilibrium overlap: its use releases one from the careful
consideration of possible crystalline or quasi-crystalline
configurations. Nevertheless, the quantity $P_{quen}(q,T)$ does not
provide any evidence for the existence or absence of RSB when avoiding
the hard task of thermalizing super-cooled liquid down to low
temperatures. One should also note that in 1RSB models both $P(q,T)$
and $P_{quen}(q,T)$ are expected to be trivial also in the region
$T_{MCT}>T>T_K$ (apart from finite size effects), just because of the
very large number $\sim e^{N \Sigma}$ of `valleys' (and corresponding
IS) which are $almost~all$ orthogonal \cite{CaGiPa}, with zero overlap
in the thermodynamic limit.

\begin{figure}[htbp]
\begin{center}
\leavevmode
\epsfig{figure=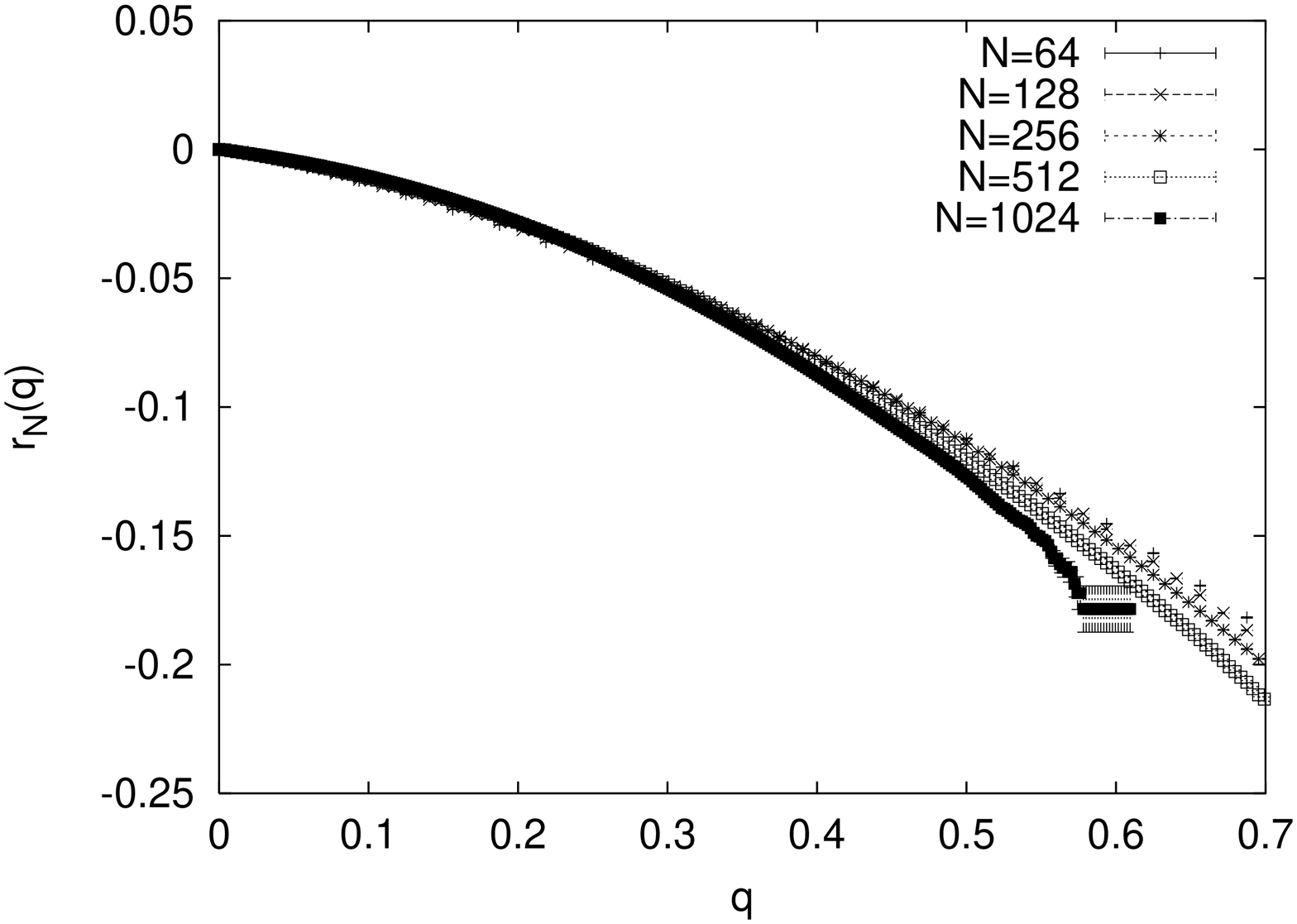,angle=270,width=5.3cm}
\epsfig{figure=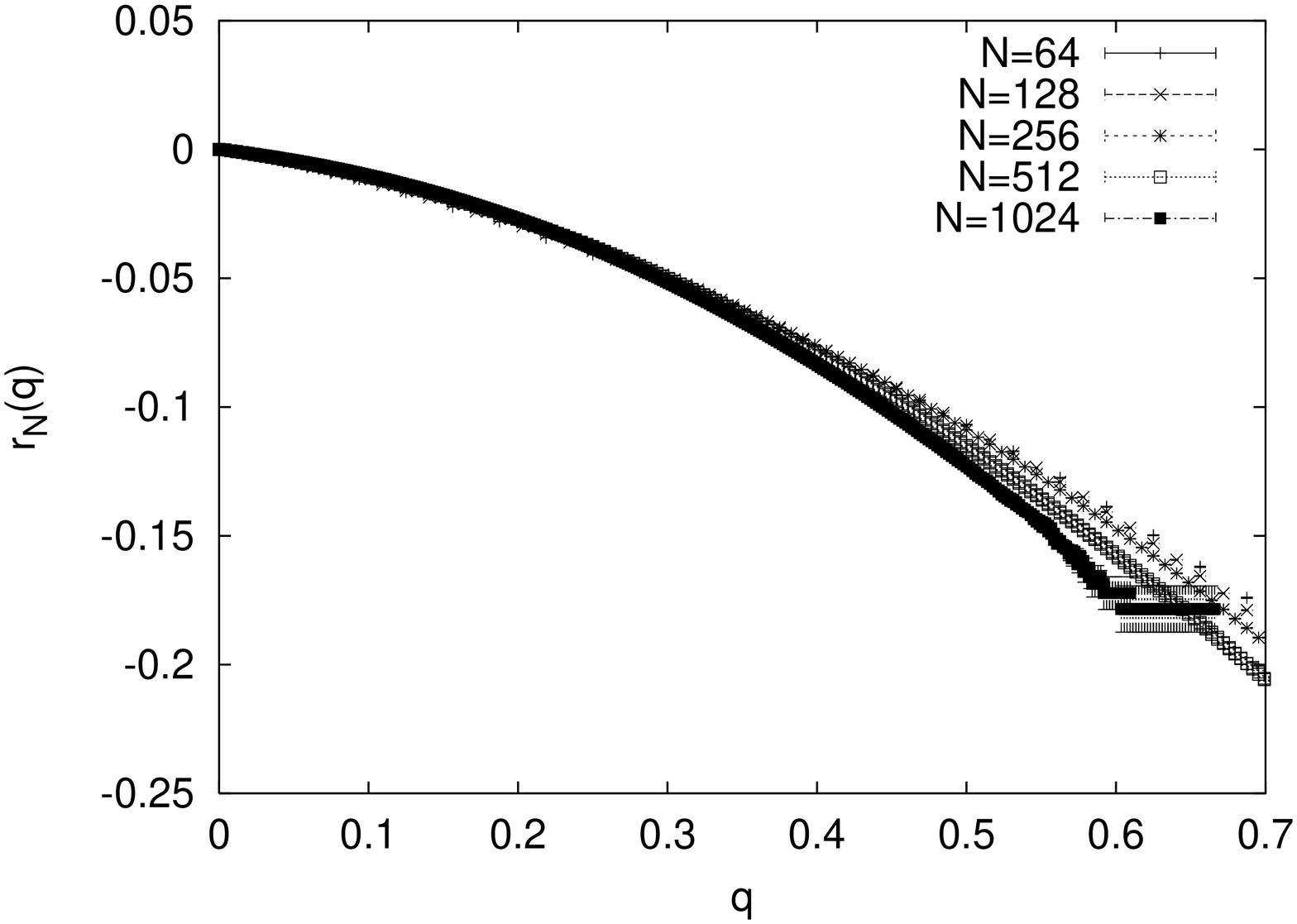,angle=270,width=5.3cm}
\epsfig{figure=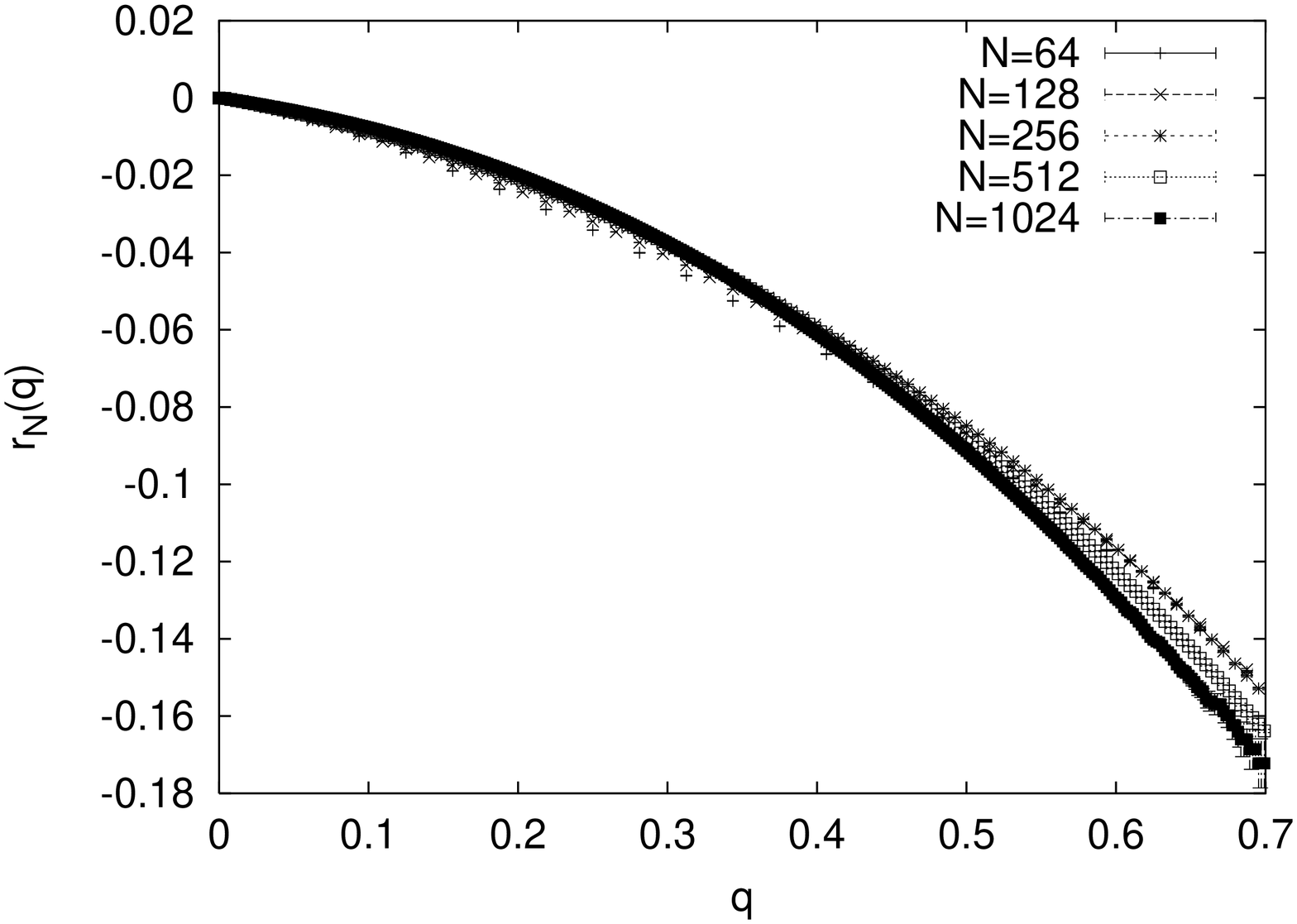,angle=270,width=5.3cm}
\caption{$r_N(q)$ for the different system sizes at the highest considered
temperature $T=3$ (left), at $T=1.8$ (center) and at $T=1.2$ close to 
$T_C$ (right).
The data plotted by using our best numerical estimate $\nu=0.68 \pm 0.05$ 
display a small dependence on $N$ in all the paramagnetic phase.}
\label{ournu}
\end{center}
\end{figure}

After clarifying this point, we note that a more careful analysis of
the data shown in [Fig. \ref{pqtc} (right)] suggests the presence of a
$weak$ RSB, as already observed in $P_{quen}(q,T=\infty)$ \cite{Pa}.
In the whole paramagnetic phase, the `quenched' probability
distribution of the overlap approaches its delta function limit
$\delta(q)$ for $N \rightarrow \infty$ much slower than the
equilibrium one. To quantify the $N$-dependence, we introduce the
function
\be
f(q) \equiv -\lim_{N \rightarrow \infty} {1 \over N} \ln \left [ P_{N,quen}(q) 
\right ].
\ee
The replica symmetry is broken $in~a~weak~sense$ if $f(q)$ is zero in
an extended region $I$ though $P_{quen}(q)$ is a $\delta$-function in the 
thermodynamic limit. This implies that $P_{N,quen}(q)$ is going to zero slower 
than exponentially in this region and that therefore, by adding to the 
Hamiltonian a quantity of order $N$ (for instance 
by using appropriate boundary conditions), one could obtain any given value of the
`quenched' overlap $q \in I$. Our best numerical evidence for such a behavior 
comes from the study of
\be
r_{N}(q,T) \equiv {1 \over N^{\nu}} \ln \left ( \int_q^1 d\:q' P_{N,quen} 
(q',T) \right ).
\ee
We get a practically $T$-independent estimate of the exponent for $T >
T_C$, $\nu=0.68 \pm 0.05$. This value is compatible with $\nu \simeq
2/3$ obtained from $T=\infty$ data in \cite{Pa}.  Though one observes
some deviations at large $q$, the weak $N$-dependence of $r_N$ shown
in [Fig. \ref{ournu}] and the value of $\nu$ that is significantly
smaller than 1, strongly suggest that $f(q)$ is zero on a finite
interval $I$, where possibly $I = [0,1]$, i.e. $P_{quen}(q)$ displays a
$weak$ breaking of replica symmetry.

\begin{figure}[htbp]
\begin{center}
\leavevmode
\epsfig{figure=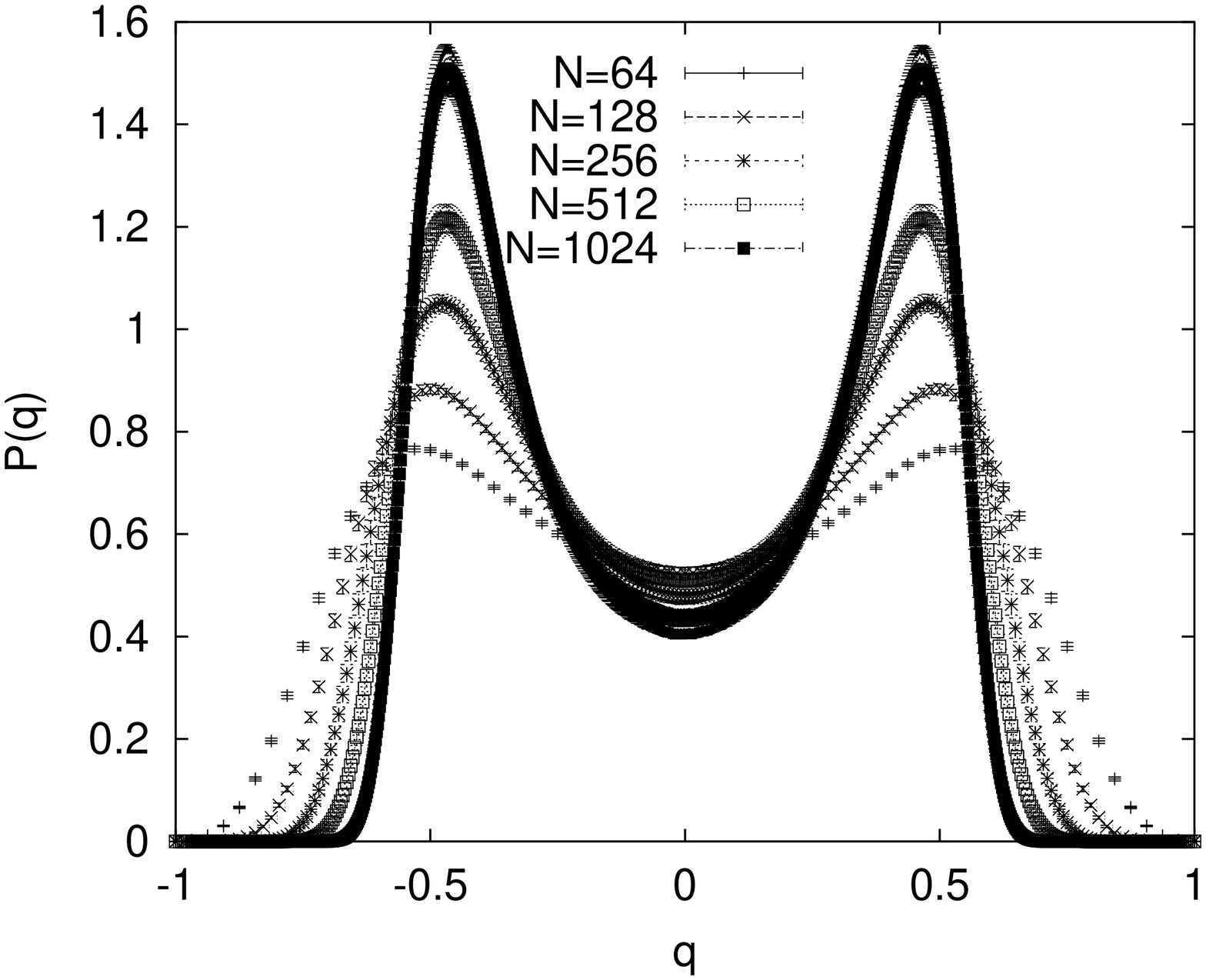,angle=270,width=8cm}
\epsfig{figure=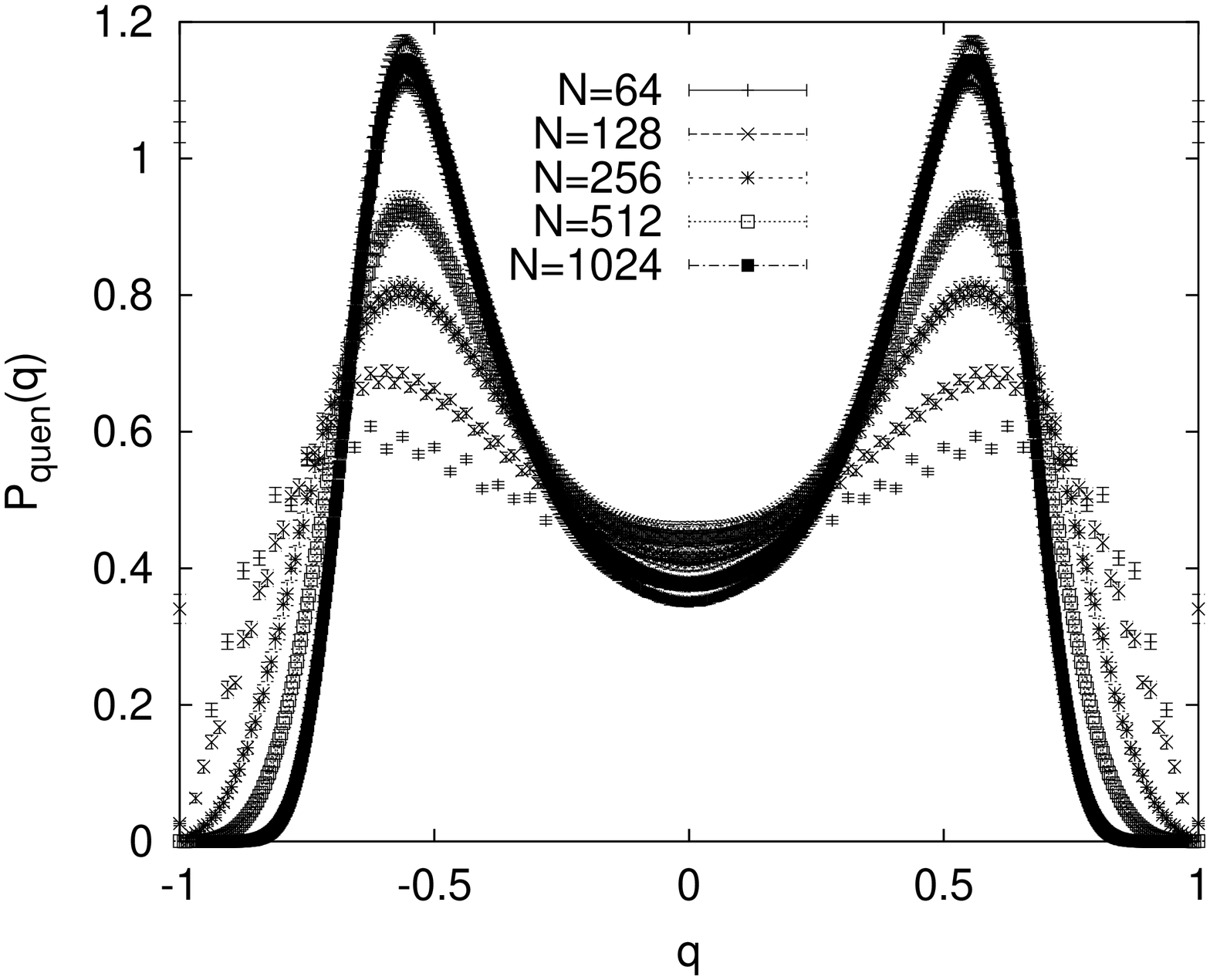,angle=270,width=8cm}
\caption{The equilibrium $P(q)$ (left) and the quenched one (right) at 
$T=0.65=0.65~T_C$ (the lowest $T$-value we have considered) 
for different system sizes.}
\label{pqglas}
\end{center}
\end{figure}

On the other hand, at temperatures lower than $T_C$, $P_{quen}(q,T)$
clearly shows the characteristic behavior corresponding to a full
replica symmetry breaking. This is what one would expect since RSB was
already evident in the $P(q)$ of the configurations we were starting
from. The qualitative similarities between $P(q)$ and $P_{quen}(q)$
are remarkable (see [Fig. \ref{pqglas}]). They are present for each
disordered configurations: for instance the number of peaks found in a
one-sample equilibrium $P_J(q,T)$ at a given $T$ is preserved in the
corresponding $P_{J,quen}(q,T)$ too. FRSB features are even more
evident when looking at $P_{quen}(q)$: in particular the presence of a
continuous $plateau$ between the two self-overlap peaks is clearer in
[Fig. \ref{pqglas}] on the right than in [Fig. \ref{pqglas}] on the
left, since the `quenched' self overlap takes larger values, though it
goes to one only for $T \rightarrow 0$.

\begin{figure}[htbp]
\begin{center}
\leavevmode
\epsfig{figure=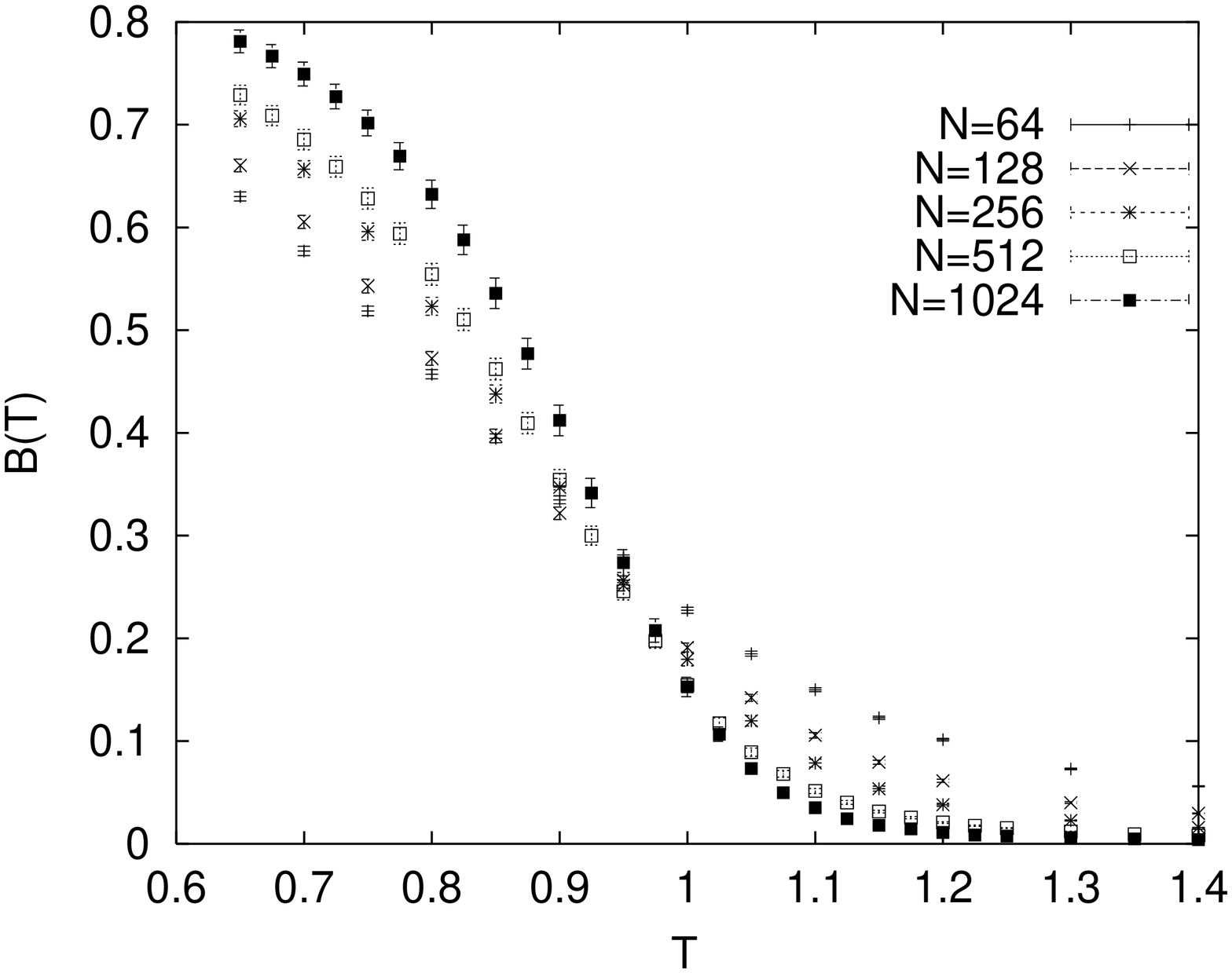,angle=270,width=8cm}
\epsfig{figure=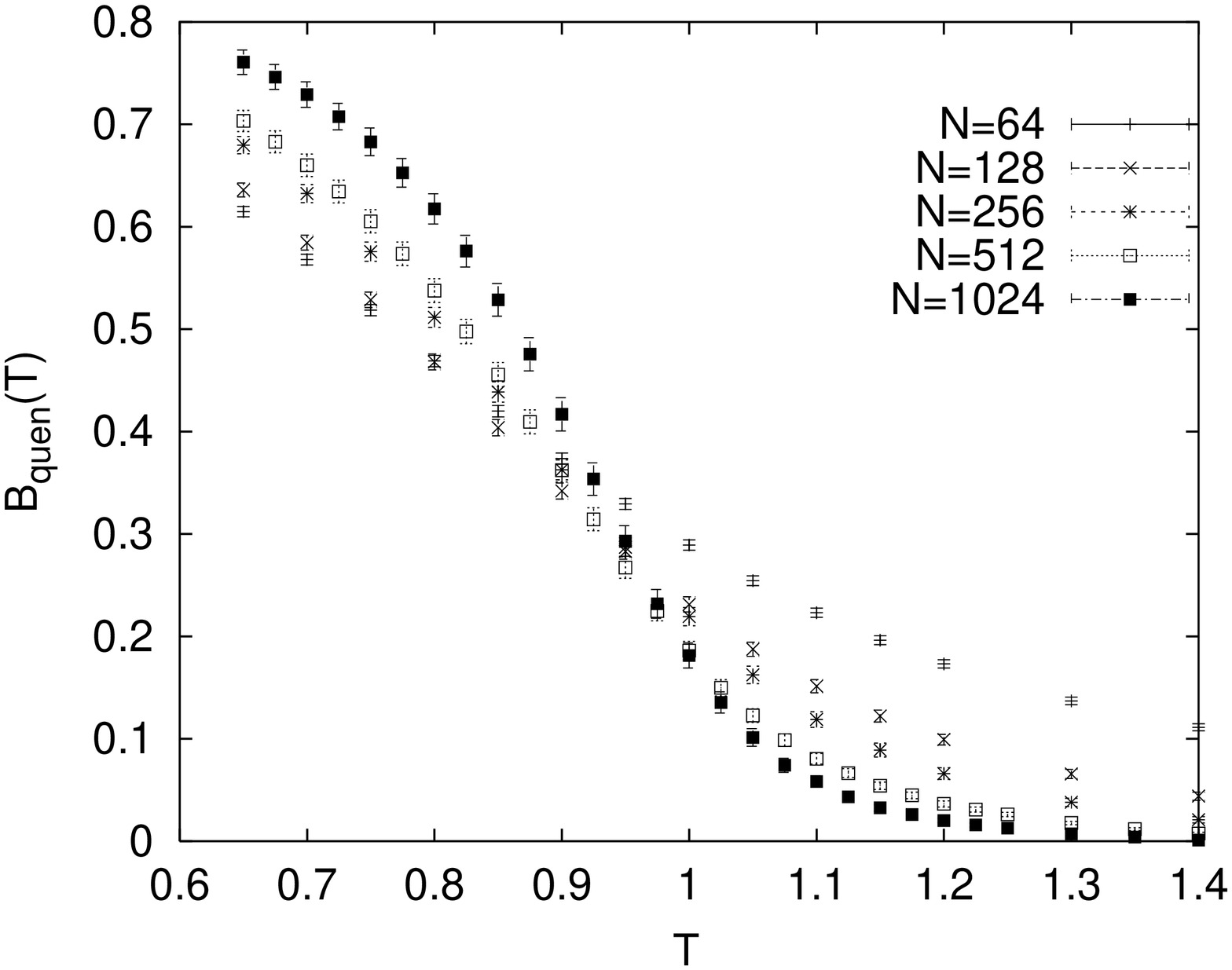,angle=270,width=8cm}
\caption{The Binder parameter from the equilibrium $P(q)$ (left) and 
from the quenched one (right) as a function of temperature for 
different system sizes.}
\label{binder}
\end{center}
\end{figure}

\begin{figure}[htbp]
\begin{center}
\leavevmode
\epsfig{figure=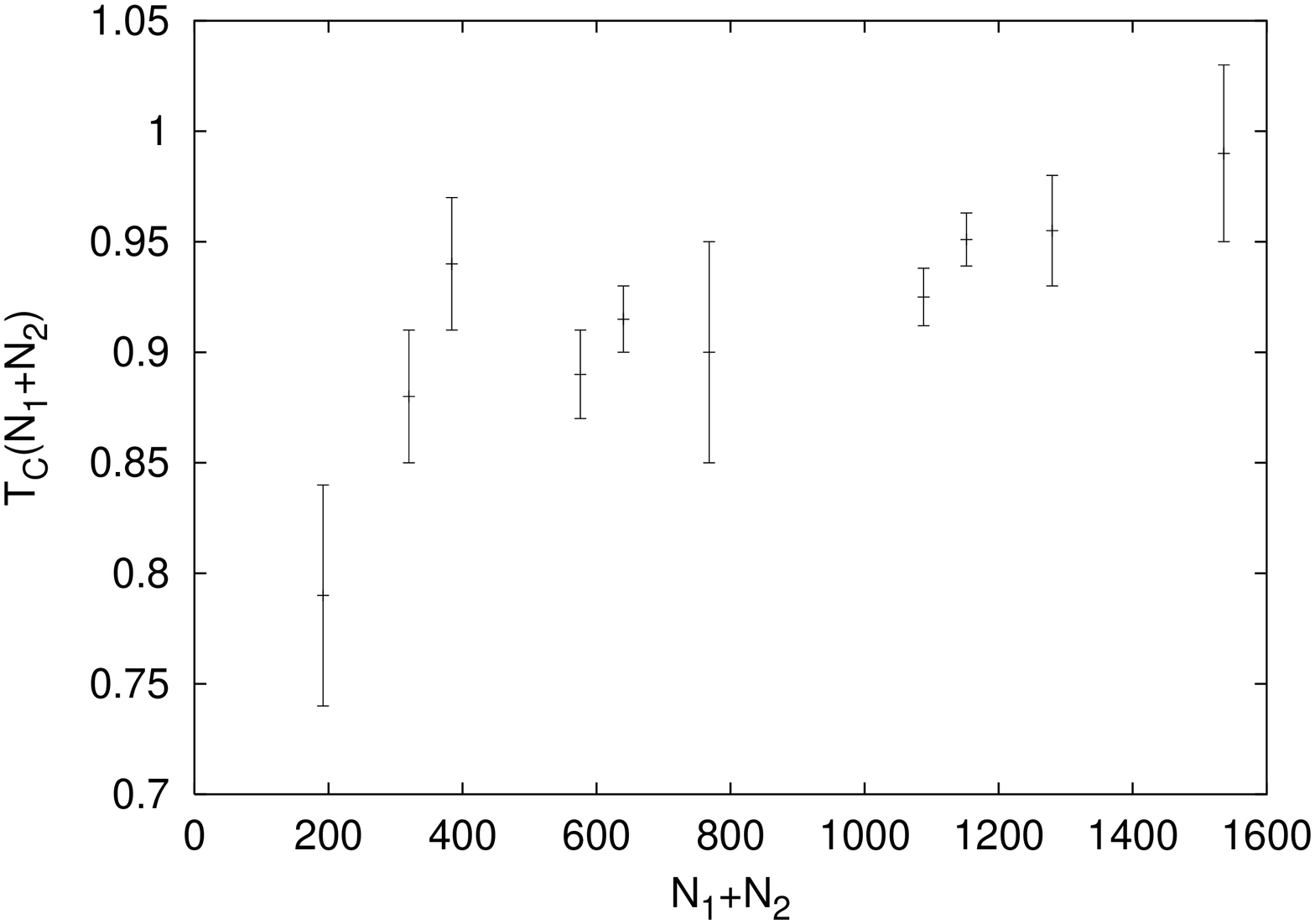,angle=270,width=10cm}
\caption{The intersection temperature of the quenched $B_{quen}(T)$ for the
different pairs $N_1$, $N_2$ of considered sizes as a function of $N_1+N_2$.}
\label{tc}
\end{center}
\end{figure}

Finite size effects play a similar role in the equilibrium
and quenched cases: for instance, it is clear from [Fig. \ref{pqglas}] that 
also the `quenched' zero overlap probability $P_{quen}(0,T)$ does not depend 
on $N$ in the glassy phase, this being a well known test when looking for FRSB 
\cite{MaPaRiRuZu}. To further investigate this point, we plot in [Fig. 
\ref{binder}] the ratio of cumulants,
\be
B(T)={1 \over 2} \left ( 3 - {\overline{\langle q^4 \rangle} \over
\overline{\langle q^2 \rangle}^2} \right ),
\ee
as a function of $T$. We recall that this is the usual quantity one
calculates in order to locate the critical temperature, since finite
size scaling predicts that curves for different sizes intersect at
$T_C$: this is the behavior observed in [Fig. \ref{binder} (left)]
(apart from corrections to scaling when considering the smaller system
sizes). Surprisingly enough, we find that the same kind of finite size
analysis can be performed on cumulants obtained from probability
distribution among IS [Fig. \ref{binder} (right)]. Though corrections
to scaling are slightly more important in this case, we see from [Fig.
\ref{tc}], where the intersection points $T_C(N_1,N_2)$ for the
different pairs $N_1$, $N_2$ of considered system sizes as a function
of $N_1+N_2$ are plotted, that one gets the correct $T_C=1$ for
$N_1,N_2 \rightarrow \infty$.

We conclude that $P_{quen}(q)$ is an interesting quantity to study.
The results we have shown could be particularly relevant when looking at 
glass-forming liquids, but the observed behavior suggests that this quantity
could also help in further clarifying the glassy phase properties of 
finite dimensional realistic spin glasses (this is a very long-standing 
subject, see for instance \cite{MaPaRiRuZu,PaYo}). 

\subsection{Correlations between equilibrium configurations and IS}

After noting the similarities between $P(q)$ and $P_{quen}(q)$
one expects the presence of strong correlations between equilibrium 
configurations and corresponding IS, particularly in the low-temperature 
phase. To quantify them, we measure the probability distribution of the 
overlaps $q_{qt}$ between each energy minimum and the configuration from 
which it is obtained, $P_{qt}(q,T)$. 

\begin{figure}[htbp]
\begin{center}
\leavevmode
\epsfig{figure=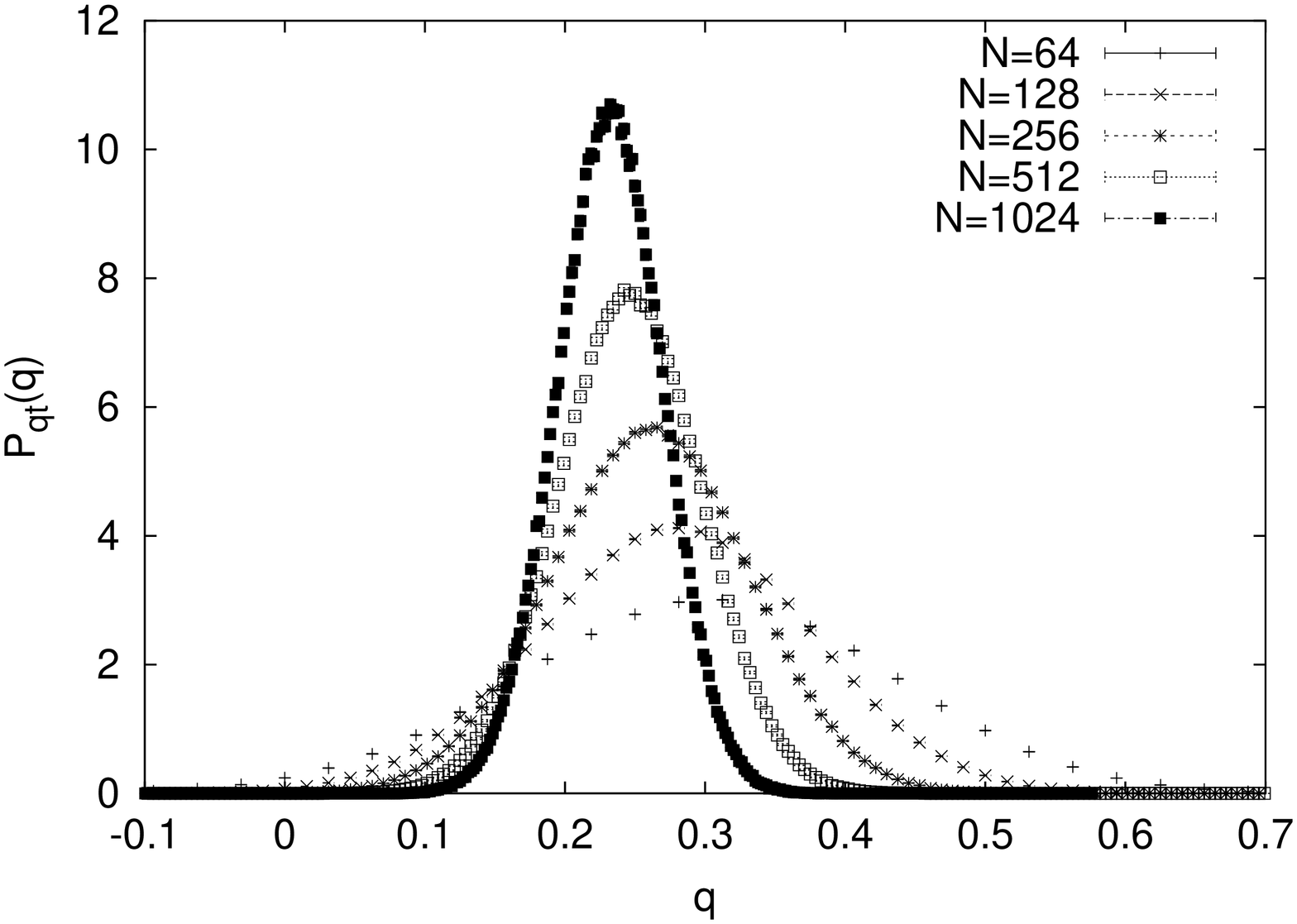,angle=270,width=5.3cm}
\epsfig{figure=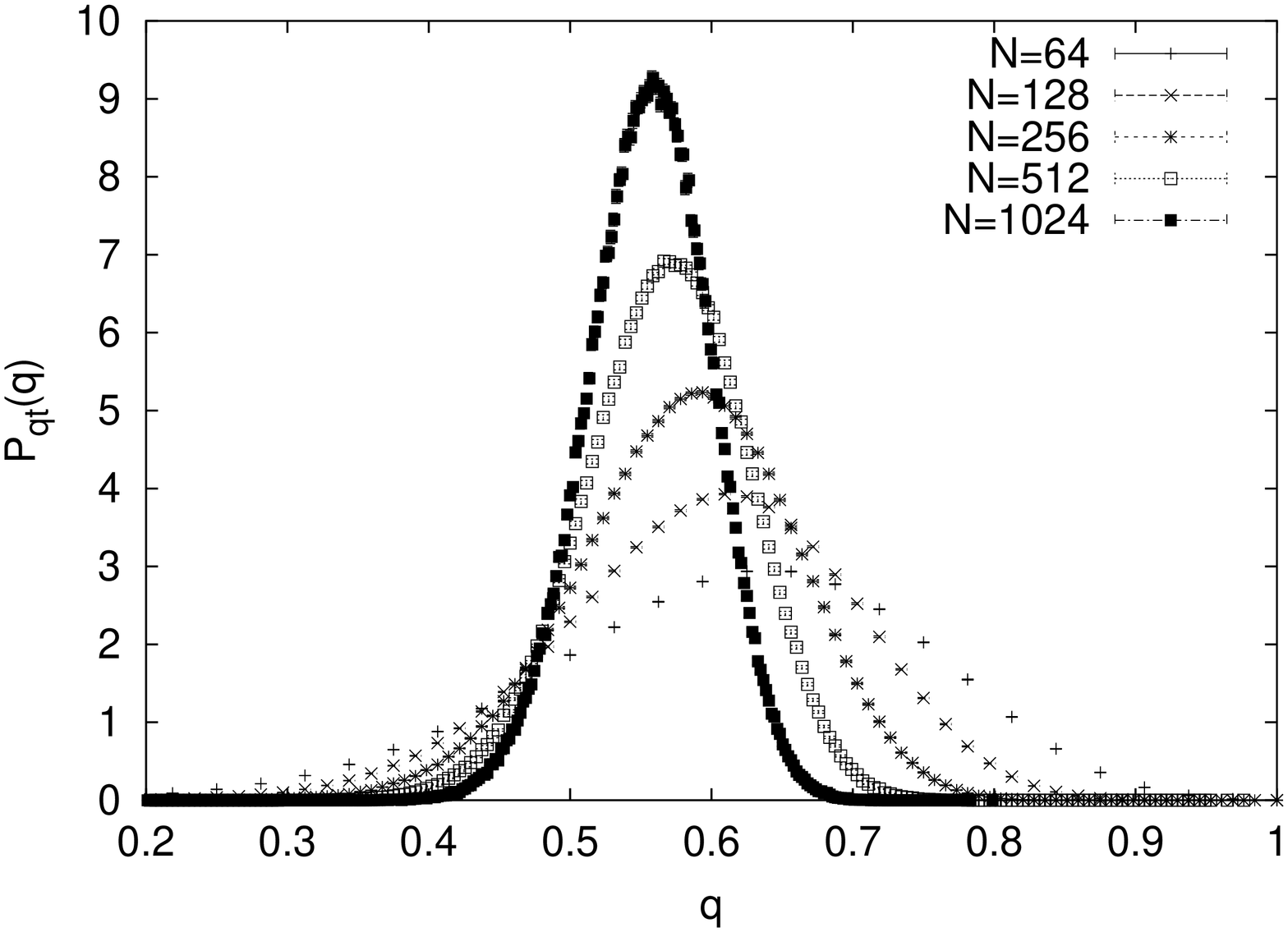,angle=270,width=5.3cm}
\epsfig{figure=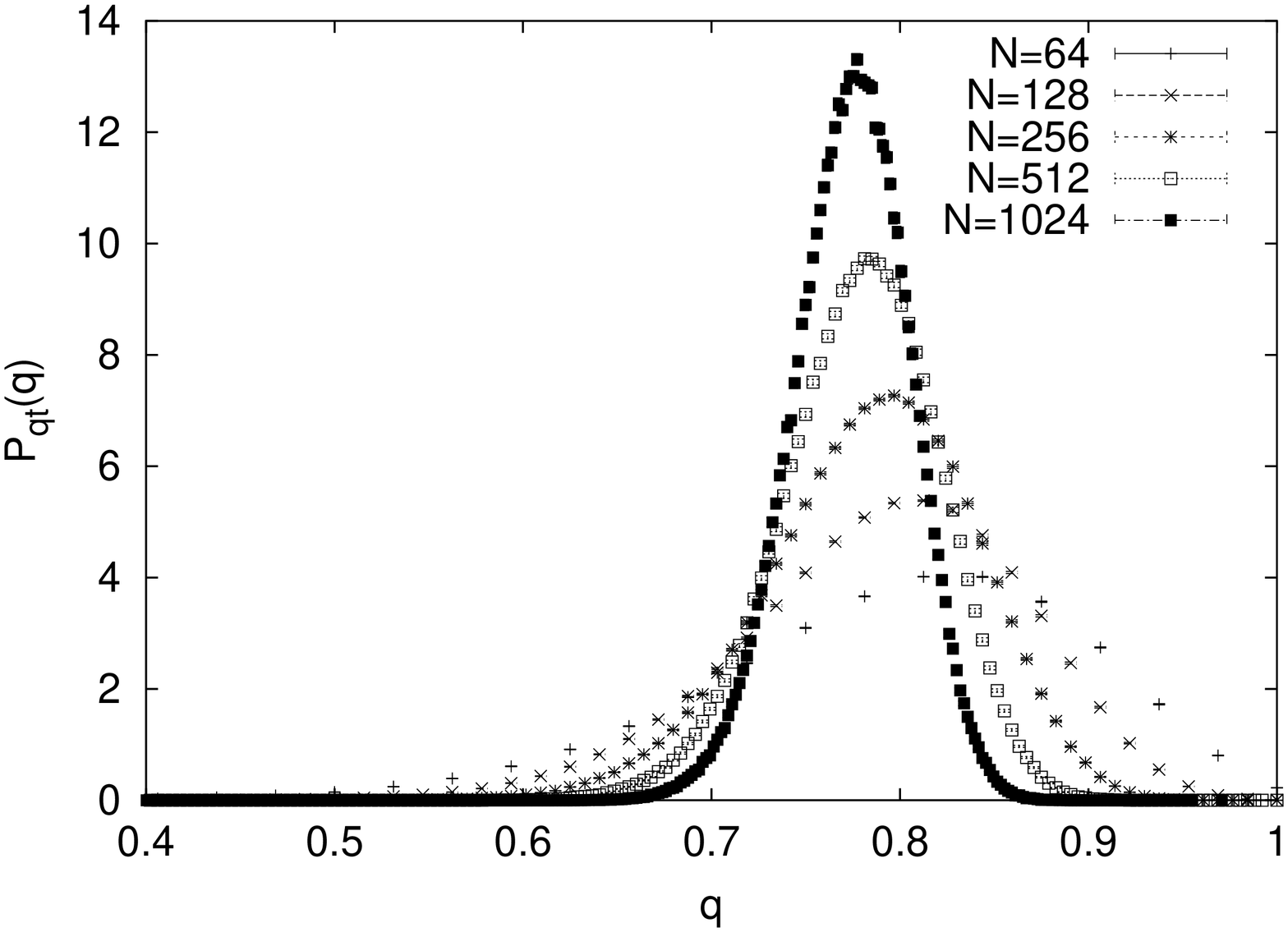,angle=270,width=5.3cm}
\caption{The probability distribution of the overlap between equilibrium 
configuration and corresponding IS $P_{qt}(q,T)$ for different sizes in the 
high temperature region at $T=3$ (left), at $T=T_C=1$ (center) and in the 
glassy phase at $T=0.65$ (right).}
\label{pqt}
\end{center}
\end{figure}

As it is shown in [Fig. \ref{pqt}], we find a Gaussian shaped distribution 
which goes toward a $\delta$-function in the thermodynamic limit both in the 
paramagnetic and in the glassy phase. We observe that below $T_C$,
at variance with $P(q)$ and $P_{quen}(q)$, this probability distribution is
a self-averaging quantity, which is easily understandable since we are 
substantially looking at overlaps between configurations related to the 
same state.

\begin{figure}[htbp]
\begin{center}
\leavevmode
\epsfig{figure=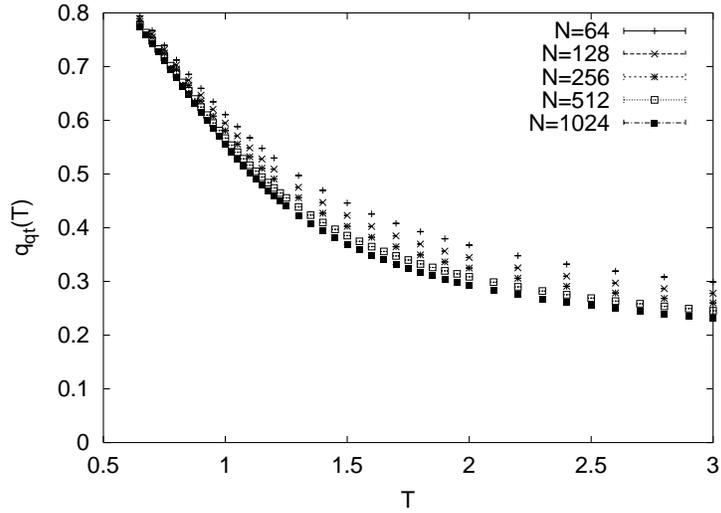,angle=270,width=10cm}
\caption{$q_{qt}(T)$ for different sizes as a function of temperature.}
\label{qqt}
\end{center}
\end{figure}

\begin{figure}[htbp]
\begin{center}
\leavevmode
\epsfig{figure=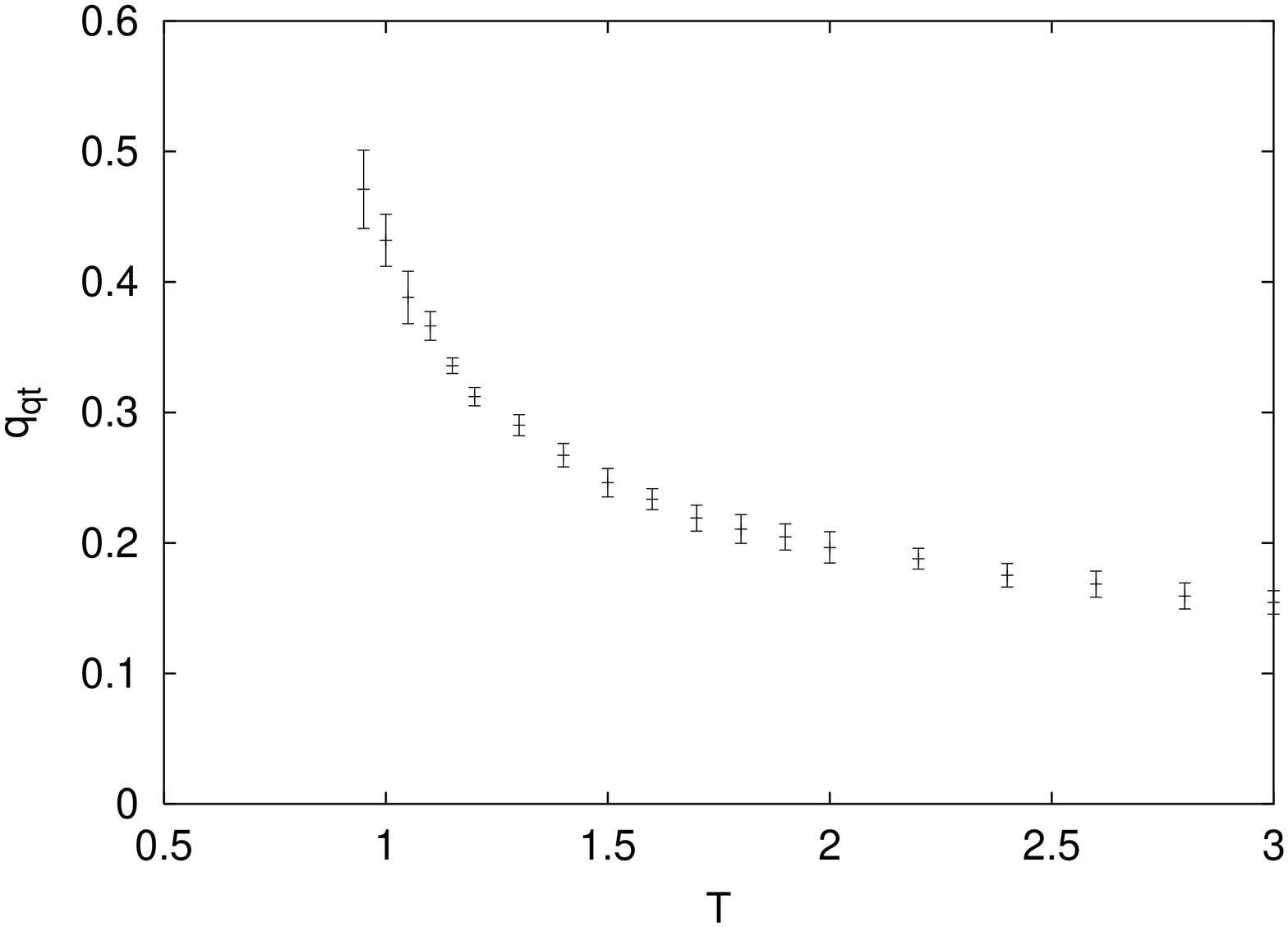,angle=270,width=8cm}
\epsfig{figure=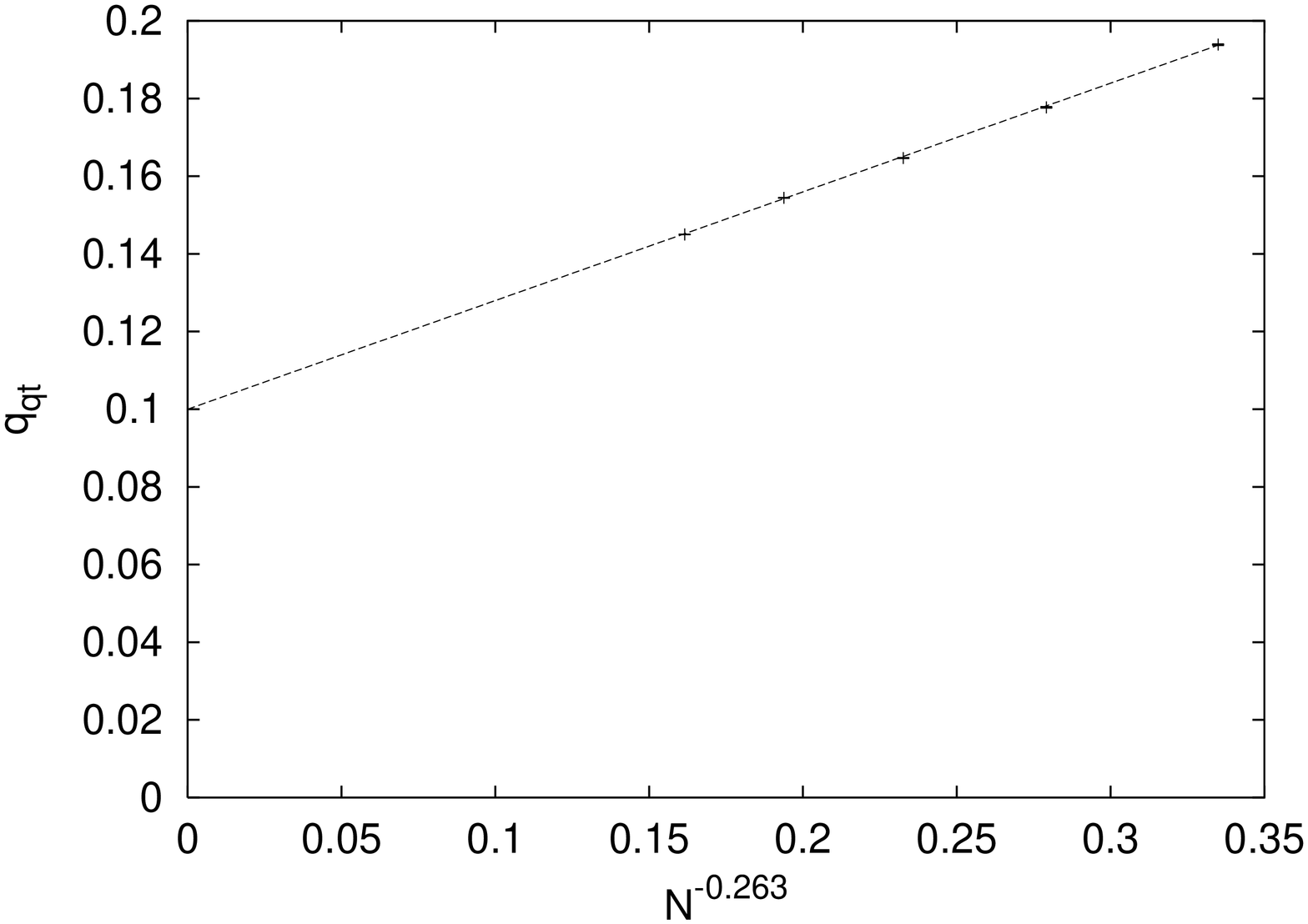,angle=270,width=8cm}
\caption{{$q_{\infty,qt}(T)$ as obtained by fitting data for different system
sizes (left) and our best fit by using 
data on the overlap between random initial conditions and corresponding 
found minima (right), which give $\lim_{T \rightarrow \infty} q_{\infty,qt}(T) 
\simeq 0.1$.}}
\label{qqtasint}
\end{center}
\end{figure}

It is intriguing to note that there is no clear evidence for the underlying 
phase transition when looking at this quantity. The mean value $q_{qt}(T)$ 
(see [Fig. \ref{qqt}]) is increasing when decreasing the temperature, and  
$\lim_{T \rightarrow 0} q_{qt}(T)=1$. The $N$-dependence appears
more pronounced in the high temperature region, where data are well fitted
by the power law
\be
q_{N,qt}=q_{\infty,qt}+{C \over N^{\alpha}}.
\label{law}
\ee 
We show in [Fig. \ref{qqtasint}] our estimates for $q_{\infty,qt}(T)$
(left), which gives in particular $q_{\infty,qt}(T_C) \sim 0.4$, and
our best fit using data for the overlap between random initial
conditions and corresponding minima (right), i.e.  $\lim_{T
  \rightarrow \infty} q_{qt}(T)$. Also in this case, we get a non-zero
value ($\sim 0.1$) in the thermodynamic limit. Nevertheless, it should
be stressed that we are confined to a relatively small range of 
system sizes, which makes a reliable estimation of the error of
$q_{\infty,qt}$ very hard.

It would be interesting to understand how much these correlations vary
when looking at different models. In agreement with our results, 
in \cite{CrRi}, for the considered volumes it was quoted 
$q_{qt} \sim 0.4$ for the SK slightly above $T_C$, to be compared with 
the higher value $q_{qt} \sim 0.94$ 
found for the 1RSB model (ROM) at a temperature higher than
the Mode Coupling $T_{MCT}$, which suggests the presence of stronger 
correlations between equilibrium configurations and corresponding IS 
in the glass-forming liquid case, and therefore a $P_{quen}(q,T)$ with a 
behavior even closer to the one at the equilibrium.

\subsection{The IS energy}

\begin{figure}[htbp]
\begin{center}
\leavevmode
\epsfig{figure=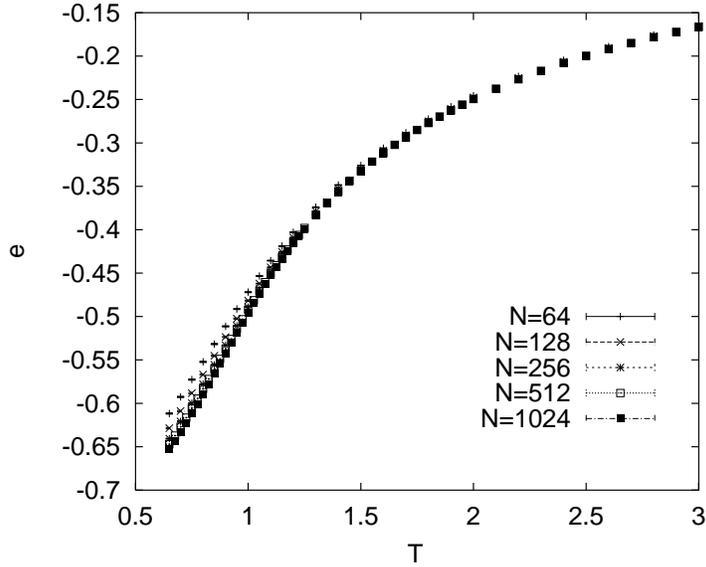,angle=270,width=10cm}
\caption{The equilibrium energy as a function 
of $T$ for the different system sizes.}
\label{ene}
\end{center}
\end{figure}

\begin{figure}[htbp]
\begin{center}
\leavevmode
\epsfig{figure=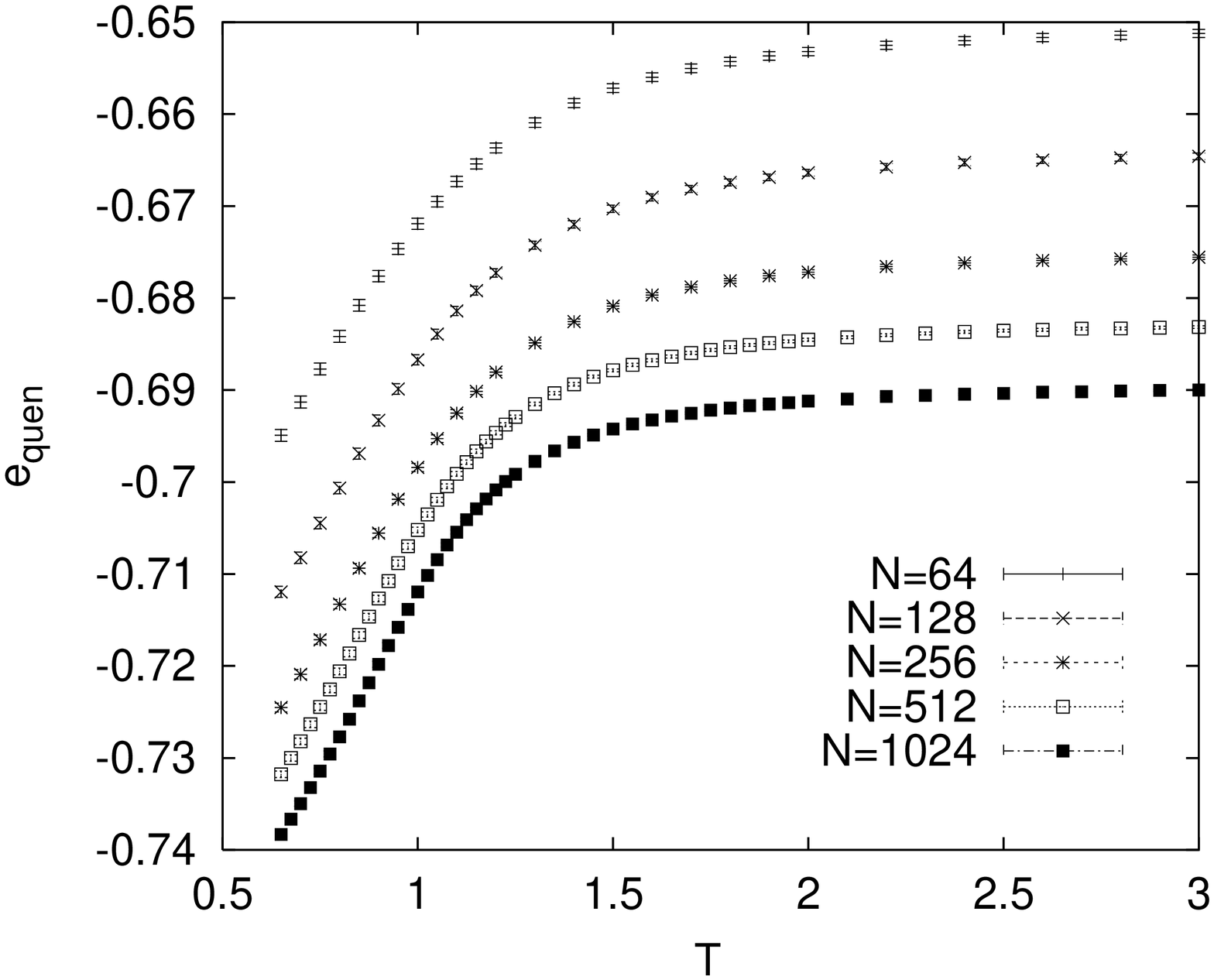,angle=270,width=7.8cm}
\epsfig{figure=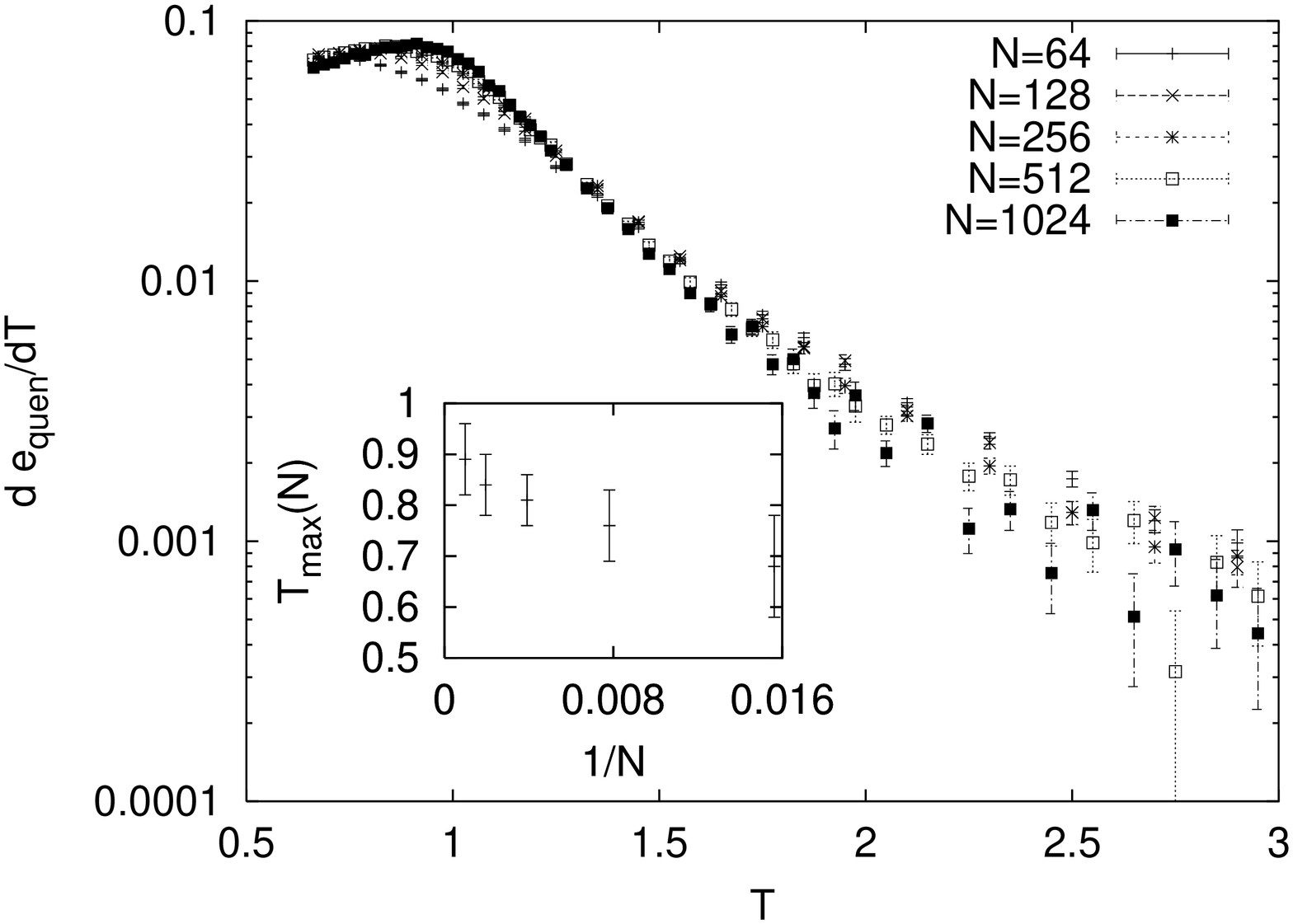,angle=270,width=8.3cm}
\caption{The quenched energy (left) and its derivative respect to the 
temperature (right) as a function of $T$ for the different system sizes $N$.
In the insert on the right, the temperature corresponding to the maximum
of $de_{quen}/dT$ is plotted as a function of $1/N$.}
\label{dene}
\end{center}
\end{figure}

In [Fig. \ref{ene}] we show the equilibrium energy $e(T)$, as a
function of the temperature $T$ (only weakly depending on the system
size).  In [Fig. \ref{dene} (left)] we present data on the IS energy
$e_{quen}(T)$, i.e. the mean energy of the minima accessible from
equilibrium configurations at a given temperature and weighted with
the Boltzmann factor of the corresponding basin. The behavior of this
quantity changes abruptly from a nearly $T$-independent value (in the
high $T$ regime) to the approximately $\propto T$ decreasing of the
low temperature region, where the IS energy continuously goes towards
the ground-state value (which is analytically known, $e_0=-0.7633$).
Correspondingly the derivative $d e_{quen}(T)/dT$, which is plotted in
[Fig. \ref{dene} (right)], displays a maximum and takes very small
high-$T$ values (note the logarithmic scale).

Data on the position $T_{max}(N)$ of the maximum of $d e_{quen}(T)/dT$ as a 
function of $N$ are shown in the insert and give evidence for 
$\lim_{N \rightarrow \infty} T_{max}(N)=1=T_C$. Our finite size analysis
therefore confirms \cite{CrRi} that in the $N \rightarrow \infty$ limit 
$e_{quen}(T)$ takes the constant threshold value \cite{CuKu2} $e_{th}$ for 
$T>T_C$. The behavior in this region agrees well with the power law 
\be
e_N=e_{th}+C/N^{\alpha},
\label{ener}
\ee
giving a constant $e_{th}=-0.7145 \pm 0.004$, which is our best numerical 
estimates for the threshold energy, down to $T \sim 1.1$ near to the critical 
temperature (and still compatible with this value also at $T_C$). 
We note that this estimates is in perfect agreement with the 
$e_{th} \simeq -0.715$ quoted in \cite{Pa} that was obtained by fitting data 
on IS reached from random initial conditions by a sequential quenching 
procedure (this value could depend
on the considered zero-temperature dynamics). The exponent $\alpha$ is slightly
increasing when going to lower temperatures and varies between 
$\alpha=0.34 \pm 0.04$ at $T=3$ and $\alpha = 0.43 \pm 0.06$ at $T=T_C$.

It is interesting to stress that finite size corrections to the  
asymptotic behavior look $very$ important, as shown by the small $\alpha$ 
value we have found. Correspondingly for all the considered sizes (up to the 
quite large volume $N=1024$) the IS energy becomes roughly constant only 
at temperatures definitely higher than $T_C$. A similar behavior with strong
finite size corrections is found in \cite{CrRi} for the considered 1RSB model 
at $T > T_{MCT}$ too. 

\section*{Conclusions}
We have presented numerical results about the properties of
energy minima in the SK model. The probability distribution of the
overlap between IS weighted with the Boltzmann factor of the
corresponding basin at temperature $T$, $P_{quen}(q,T)$, turns out to be
qualitatively similar to the equilibrium overlap distribution $P(q,T)$
at the same temperature $T$. We found a trivially-shaped $P_{quen}(q,T)$
in the whole paramagnetic ($T > T_C=1$) phase, whereas the FRSB
behavior characteristic of the glassy phase is evident from data on
$P_{quen}(q,T)$ only when looking at energy minima obtained from
equilibrium configurations at temperature definitely lower than $T_C$.
A finite size analysis of the Binder parameter for $P_{quen}(q,T)$
gives the same estimate of the critical temperature as the
usual (equilibrium) one.

These results can be particularly relevant for glass-forming liquids,
where the overlap is more precisely definable between IS
\cite{BhBrKrZi}, but they also imply that numerical evidence for
replica symmetry breaking can not be extracted from data on
$P_{quen}(q,T)$, which have been obtained from configurations at
equilibrium at temperatures {\it above} the possible glass
transition temperature (in that case $T_K < T_{MCT}$).

The analysis of finite size corrections to the IS energy confirms
\cite{CrRi} that it is approaching the expected thermodynamic limit
behavior $e_{IS}=const=e_{th}$ for $T>T_C$, whereas it appears
continuously decreasing for $T<T_C$.

\section*{Acknowledgments}
B.C. acknowledges Uwe M\"ussel and Holger Wahlen for useful suggestions, and 
would like to thank the John-von-Neumann Institut f\"ur Computing 
(Forschungszentrum J\"ulich), where this work was partially developed.
Simulations were ran on the Forschungszentrum J\"ulich Cray T3E.

\end{document}